\documentclass[aps,pre,reprint,onecolumn,superscriptaddress,nofootinbib,longbibliography,floatfix]{revtex4-2}

\usepackage{amsmath,amssymb}
\usepackage{graphicx}
\usepackage{booktabs}
\usepackage{lmodern}
\usepackage[T1]{fontenc}
\usepackage[colorlinks=true,linkcolor=blue,citecolor=blue,urlcolor=blue]{hyperref}

\begin{document}

\title{Why wait? Optimal rules for intertemporal decisions across life history, economics and machine learning}

\author{Elias Tannira}
\email{tannirae@uchicago.edu}
\affiliation{Department of Ecology and Evolution, University of Chicago, Chicago, IL 60637, USA}
\affiliation{Santa Fe Institute, 1399 Hyde Park Road, Santa Fe, NM 87501, USA}

\author{Lu\'is M. A. Bettencourt}
\affiliation{Department of Ecology and Evolution, University of Chicago, Chicago, IL 60637, USA}
\affiliation{Santa Fe Institute, 1399 Hyde Park Road, Santa Fe, NM 87501, USA}
\affiliation{Data Science Institute, University of Chicago, Chicago, IL 60637, USA}

\date{\today}

\begin{abstract}
When should we take a smaller reward now rather than wait for a larger one later? This question recurs across evolutionary biology, behavioral ecology, psychology, economics and artificial intelligence, yet each field treats temporal discounting differently. We show that intertemporal decisions across these disciplines share a common structure and a single optimal rule: wait while the growth rate of the reward exceeds the discount rate, and act when the two rates become equal. For a single decision, this rule defines an optimal waiting time; for a continuous reward stream, it becomes an equation of motion for quantities apportioned over time, such as consumption or fertility. We derive reward and discount rates from explicit mechanisms, including mortality and other hazards, uncertainty and risk aversion, population growth, capital depreciation, dread, subsistence thresholds and the value of information, with examples from laboratory choice, retirement, maturation, resource extraction, capital accumulation and exploration. The framework shows how apparent anomalies, including hyperbolic discounting, the sign effect and the impatience of deprivation, can arise from the relative structure of rewards and hazards rather than from fixed preferences. Finally, we identify what must be measured to explain and modify temporal preferences in experiments and policy.
\end{abstract}

\maketitle

\section{Introduction}

Patience is commonly seen as a virtue, yet impatience is universal. This apparent paradox is observed in laboratory experiments, everyday behavior, and policy. Children often take one marshmallow now instead of two later~\cite{mischel1989}; pigeons, rats and monkeys tend to prefer small immediate rewards to larger delayed ones, even when waiting would raise their  overall food intake~\cite{hayden2016,stephens2004}; people with low socioeconomic status tend to discount the future more steeply than the affluent~\cite{lawrance1991,reimers2009,haushofer2014}. Artificial agents trained by reinforcement learning are made impatient by design, through a discount factor that weighs each future step less than the last~\cite{sutton2018}, even when their performance improves with more exploration~\cite{jiang2024rethinking,chen2021decision}. All of these agents face the same dilemma: how long should one wait?

Even though these questions share commonalities, each discipline has developed its own approach. Economics tends to represent time preferences by discounted-utility models, in which future utility gains are weighted by a constant rate of pure time preference~\cite{samuelson1937,fisher1930interest,frederick2002}. Laboratory work has documented many systematic departures from this assumption: discount rates that fall with delay, larger rewards discounted less than small ones, losses discounted less than gains, and preferences that reverse as rewards draw near~\cite{thaler1981,loewenstein1992,frederick2002,strotz1955,laibson1997}. These ``anomalies'' have become central to the sub-field of behavioral economics, but most remain unexplained~\cite{laibson1997golden,frederick2002time}.  Psychology has come to describe some of these patterns with hyperbolic discount functions~\cite{ainslie1975,mazur1987,rachlin2006} whose fitted parameters vary with the commodity, the age of the subject and its psychiatric condition~\cite{odum2020,lu2023,story2016,bickel2001}. Evolutionary biology, instead, derives discounting from first principles through Fisher's reproductive value. Future reproduction is devalued by mortality hazard and by background population growth~\cite{fisher1930,hamilton1966,rogers1994,hansson1990}, which predicts age-specific behavioral profiles in life history theory~\cite{rogers1997,sozou2003,chu2010,kageyama2011}. Behavioral ecology has tested whether the mortality risk of losing delayed rewards explains animal impulsiveness, but this risk is often found to be too weak, suggesting additional mechanisms~\cite{kacelnik1997,stephens2004,henly2008,stevens2010}. Reinforcement learning, an important subfield of artificial intelligence (AI), treats the discount factor mostly as a hyperparameter chosen for computational convenience~\cite{sutton2018,fedus2019}. Recent developments in AI show how increasing the agent's context window, by reducing discounting, can dramatically improve performance in complex tasks~\cite{jiang2024rethinking,chen2021decision}.

These diverse fields rarely draw on each other, so their results and methods  have remained difficult to compare~\cite{frederick2002,cohen2020}. Among the open problems, measured discount rates are poorly identified and weakly predictive of behavior outside the laboratory~\cite{cohen2020,lipman2024,urminsky2015}. Mortality alone predicts discount rates one to two orders of magnitude smaller than those observed in humans and animals~\cite{mell2021,stephens2004}. As a consequence of these anomalies, each field has added its own corrections, including opportunity costs of waiting~\cite{fawcett2012,mell2021}, endogenous or state-dependent impatience~\cite{obstfeld1990,hayakawa1997,becker1997,trostel2001}, uncertainty about hazards~\cite{sozou1998,story2016,robson2009}, or minimal subsistence thresholds~\cite{decourson2023}. Several authors have called for a more unified account of these proposals~\cite{chu2010,kageyama2011,mell2021,fawcett2012}, but none, to our knowledge, has been carried out across fields. A recurrent difficulty is that discounting is studied in isolation, as a property of the agent, and by relatively opaque mathematical methods.  By contrast, a decision to wait is contextual and relatively intuitive: in real situations,  agents must always weigh discounting against how rewards and associated assets may increase over time.

Here we ask whether a common structure underlies intertemporal decisions across these fields, such that discount rates and waiting times can be derived rather than assumed. Our starting point is that every such decision involves two functions of time: a reward, $R$, which typically grows if one waits, and a discount function, $G$, which declines. Their logarithmic rates of change, $\gamma_R$ and $\gamma_G$, give a single decision rule: wait while $\gamma_R>\gamma_G$ and act when $\gamma_R=\gamma_G$. We distinguish point problems, with one decision and an optimal waiting time, from path problems, in which an agent continuously divides a stock between present use and some sort of investment. We show that path problems reduce to point problems for a single marginal unit held over time, so that the same rule becomes the Euler equation of economics, the Bellman condition of reinforcement learning, and the optimality condition of life history theory, Tables~\ref{tab:point-path} and~\ref{tab:traditions}.

Within this framework, discounting has three broad origins. The first is intrinsic impatience, assumed in much of psychology and economics but typically left unexplained. The second, and most important, is the hazard of not being able to collect the reward later: this may be due to the agent's own mortality or incapacity, environmental events, or the perishability of the reward itself. The third is the evaluation of nominal reward value against a standard that itself changes in time such as population growth in life history theory, or inflation and economic growth in economics. Other mechanisms, notably uncertainty combined with risk aversion, diminishing returns, subsistence thresholds and the value of information, act on the growth of rewards but are often observed as discounting.

The paper makes three main contributions. First, we give a common language, expressed as a single transparent decision rule, for intertemporal problems in evolution, ecology, psychology, economics and machine learning. Second, we show how discount rates and reward growth rates emerge from explicit mechanisms and, specifically, that a slower growth of rewards is observationally equivalent to faster discounting. Third, through a sequence of worked examples, we show that several well-known anomalies, such as hyperbolic discounting, the sign effect, the U-shaped profile of time preference over the life course and the ``impatience of the poor'', can be rational responses to the structure of rewards and hazards rather than subjective traits. The structure of the paper is as follows: section~\ref{sec:types} sets up the framework, sections~\ref{sec:point examples} and~\ref{sec:path examples} develop examples of point and path problems, and section~\ref{sec:discussion} discusses implications, elaborations, and open problems.

\section{General structure of intertemporal decisions}
\label{sec:types}
We now consider models for decisions whose consequences unfold over time. Often this follows the agent's life course, so that time can also be read as age. We distinguish two types of problem: point problems with a single decision, and path problems with a sequence of decisions over time. Because path problems generalize point problems, we treat them in sequence.

\subsection{Point problems}
\label{sec:point}
Point problems involve a single choice. For example, more money (or marshmallows, or food) later versus less now, or a larger loss later versus a smaller one now. These are the typical decisions studied in laboratory settings in psychology and behavioral economics~\cite{frederick2002,urminsky2015,cohen2020}.

Let $R(\tau,t)\geq 0$ be the magnitude of a reward to be collected at time $\tau\geq t$, where $t$ is the earliest possible collection time.  In interesting problems, $R$ increases with $\tau$ so that, in the absence of other considerations, it is worth waiting. A loss that grows in time is the mirror image (a negative reward) so without discounting one should never wait. Although rewards increase with time, it is assumed that the agent values them less when they are received later. This effect is temporal discounting: we represent it by a function $0<G(\tau,t)\leq 1$, with $G(t,t)=1$. Together, $R$ and $G$ define the subjective value of a reward collected at $\tau$,
\begin{align}
    \label{eq:reward contribution}
    V(\tau,t) = R(\tau,t)\,G(\tau,t),
\end{align}
which is the quantity to be maximized. When $t=0$ (e.g.\ birth), we write $R(\tau)$, $G(\tau)$ and $V(\tau)$. Reward and discounting are characterized by their temporal rates,
\begin{align}
    \gamma_R(\tau,t) \equiv \frac{\partial}{\partial \tau}\ln R(\tau,t), \qquad \gamma_G(\tau,t) \equiv -\frac{\partial}{\partial \tau}\ln G(\tau,t), \qquad \frac{\partial \ln V}{\partial \tau}= \gamma_R - \gamma_G,
    \label{eq:rates}
\end{align}
so that $G(\tau,t)=\exp[-\int_t^\tau \gamma_G(u)\,du]$, where $u$ is a dummy integration variable (time); $\gamma_G$ may depend on environmental or psychological factors. This discount rate, $\gamma_G$, is the quantity estimated in psychological and economic experiments~\cite{frederick2002,cohen2020,lipman2024}. Constant $\gamma_G$ gives exponential discounting, whereas declining $\gamma_G$ gives hyperbolic discounting, see below.

Equation~\eqref{eq:rates} gives the decision rule. If $\gamma_R<\gamma_G$, the subjective value falls over time and the reward should be collected immediately. If $\gamma_R>\gamma_G$, subjective value rises and the agent should wait. Waiting ends at the time $\tau_*$ at which $\gamma_R(\tau_*)=\gamma_G(\tau_*)$ and $\gamma_R-\gamma_G$ changes sign from positive to negative. Then, $\tau_*$ is the optimal waiting time. Point problems are simple because $R$ and $G$ are known functions of time. Laboratory experiments can create this structure by construction but real decisions are affected by uncertainty and often let the agent consume or invest partially over time. Such continuous dynamical allocations become path problems.

\subsection{Path problems}
\label{sec:path}
In path problems, the agent decides continuously in time. The solution is a temporal trajectory reflecting an optimal behavior, such as a rate of savings, consumption or (biological) investment.
Consider an agent with a resource stock $k(t)$ that grows at a rate $h(k)$ and is depleted through consumption $x(t)$,
\begin{equation}
\dot{k}=h(k)-x.
\label{eq:stock dynamics}
\end{equation}
The agent has a choice of a continuum of consumption trajectories $x(s)$, for $s\geq t$, to maximize total discounted rewards
\begin{equation}
W(t)=\int_t^\infty F[x(s)]\,G(s,t)\, ds.
\label{eq:path problem}
\end{equation}
Here, $F[x(s)]$ is the reward stream obtained from consumption at time $s$. It is discounted through time by $G(s,t)$. The stock $k(t)$ may be capital in economics, body mass in life history theory, or a natural resource in the ground. When the reward stream is concentrated at a single collection time $\tau$, such that $F[x(s)]=R(\tau,t)\delta(s-\tau)$, we recover the point problem.
The optimal trajectory is determined by measuring the consequences of transferring a small amount of consumption between two times.  Suppose the agent gives up a small amount of consumption $\varepsilon$ at time $t$ and invests it until $s>t$. The consumption unit is added to the stock and appreciates by the factor 
\begin{equation}
\Phi(s,t)=\exp\left[\int_t^s h'\big(k(u)\big)\,du\right],
\qquad \Phi(t,t)=1.
\label{eq:Phi}
\end{equation}
The variation in the consumption path resulting from the transfer of $\varepsilon$ between two times is
\begin{equation}
\delta x(u)=-\varepsilon\delta(u-t)
+\varepsilon\Phi(s,t)\delta(u-s).
\label{eq:two-pulse}
\end{equation}
To first order, the resulting change in total discounted reward is
\begin{equation}
\delta W=\varepsilon\bigg[\underbrace{
F'[x(s)]\Phi(s,t)}_{R(s,t)}G(s,t)-\underbrace{F'[x(t)]}_{R(t,t)}
\bigg]=\varepsilon\left[V(s,t)-V(t,t)\right].
\end{equation} 
The reward of one unit held is $R(s,t)=F'[x(s)]\,\Phi(s,t)$. This is  the value of cashing it in times appreciation. Because units can be moved in either direction across time, and every feasible variation is a superposition of such moves (Appendix~\ref{app:Phi}), an optimum requires $\delta W=0$ for all $s$. Thus, $V(s,t)$ stays constant as $s$ varies, rather than peaking at a single value, and
\begin{equation}
\gamma_R(s) = \frac{d\ln F'\big[x(s)\big]}{ds} + h'\big(k(s)\big) = \gamma_G(s) \quad\text{for all } s.
\label{eq:path condition}
\end{equation}
Because $x(s)$ appears in $\gamma_R$, this is an equation of motion: the optimal intertemporal behavior.

Whether we have a point or path problem depends on the curvature of $F$. If $F$ is linear in $x$ (or the reward is indivisible), $R$ does not depend on the path, all units are best cashed in at the peak of $V$, and we have a point problem. If $F$ is concave, cashing in more at a single time lowers the marginal reward, which encourages spreading $x$ over time until $V$ is flat, which makes it a path problem. Table~\ref{tab:point-path} summarizes the two types of problem. Table~\ref{tab:traditions} maps problems across different disciplines onto these shared structures. A common theme of the examples below is that higher discounting lowers total rewards, such as size at maturity or lifetime utility. Yet present-oriented behavior under high hazards, uncertainty, or background growth or depreciation is the actual rational response, not a pathology of character or an undesirable trait~\cite{kidd2013,story2016,daly2005}.

\begin{table}[t]
\centering
\small
\renewcommand{\arraystretch}{1.3}
\setlength{\tabcolsep}{4pt}
\begin{tabular}{@{}p{2.52cm}p{6.02cm}p{7.00cm}@{}}
\toprule
 & \raggedright \textbf{Point problem} & \raggedright \textbf{Path problem} \tabularnewline
\midrule
Choice & \raggedright a single collection time $\tau\geq t$ & \raggedright a trajectory $x(s)$, $s\geq t$ \tabularnewline
Objective & \raggedright $V(\tau,t)=R(\tau,t)\,G(\tau,t)$ & \raggedright $W(t)=\int_t^\infty F[x(s)]\,G(s,t)\,ds$, subject to $\dot k=h(k)-x$ \tabularnewline
Unit held & \raggedright the reward itself & \raggedright one marginal unit of $x$, withheld at $t$ and cashed in at $s$ \tabularnewline
Reward $R$ & \raggedright a given function of $\tau$ & \raggedright $F'[x(s)]\,\Phi(s,t)$: value of cashing in $\times$ appreciation \tabularnewline
Optimality & \raggedright $\gamma_R(\tau_*)=\gamma_G(\tau_*)$: $V$ peaks & \raggedright $\gamma_R(s)=\gamma_G(s)$ for all $s$: $V$ flat, $\delta W=0$ \tabularnewline
Result & \raggedright optimal waiting time $\tau_*$ & \raggedright equation of motion for $x(s)$ \tabularnewline
Arises when & \raggedright indivisible reward, \\ or $F(x) \propto x$ & \raggedright divisible stock with diminishing returns ($F''<0$) \tabularnewline
Limits & \raggedright $\gamma_R>\gamma_G$ always: wait forever; $\gamma_R<\gamma_G$ always: collect now & \raggedright where constraints bind (e.g.\ subsistence), $\gamma_R\neq\gamma_G$ \tabularnewline
\bottomrule
\end{tabular}
\caption{Two types of intertemporal decision problems. The decision always compares the growth rate of the reward, $\gamma_R$, with the discount rate, $\gamma_G$, Eq.~\eqref{eq:rates}. A point problem is a path problem whose reward flow is concentrated at a single time. The appreciation factor $\Phi$ follows from the stock dynamics, Eq.~\eqref{eq:Phi}.}
\label{tab:point-path}
\end{table}

\begin{table}[t]
\centering
\footnotesize
\renewcommand{\arraystretch}{1.35}
\setlength{\tabcolsep}{3pt}
\begin{tabular}{@{}p{2.45cm}p{3.22cm}p{3.22cm}p{3.50cm}p{3.71cm}@{}}
\toprule
\raggedright\textbf{Tradition} & \raggedright\textbf{Objective} & \raggedright\textbf{Discount rate $\gamma_G$ (sources)} & \raggedright\textbf{Point problem $\gamma_R=\gamma_G$} & \raggedright\textbf{Path problem $\gamma_R=\gamma_G$} \tabularnewline
\midrule
\raggedright Psychology,  behavioral economics
& \raggedright $V=R\,G$ for a reward (loss)
& \raggedright impatience; hazard; dread, savoring
& \raggedright linear reward, $R\propto\tau$: $1/\tau_*=\gamma_G$; \\ losses: $\gamma_L=\gamma_G$; \\ noisy rewards: $g-\eta\frac{\sigma^2}{2}=\gamma_G$
& \raggedright rarely posed  \tabularnewline
\raggedright Economics
& \raggedright $\int u(c)\,e^{-\gamma_G t}\,dt$, subject to capital growth
& \raggedright pure time preference; \\ risk, $\eta\sigma^2/2$; subsistence
& \raggedright retirement timing 
& \raggedright Hotelling: $\dot p/(p-c_e)=\gamma_G$; RCK: $f'-\delta_k -\eta\, \frac{\dot c}{c}=\gamma_G$ \tabularnewline
\raggedright Life history theory
& \raggedright maximize $r$, subject to Euler--Lotka eq.
& \raggedright $r+\mu$: population growth and mortality hazard
& \raggedright age at maturity: $m'/m=r+\mu_J$; linear $\phi$: $P'(w_*)=r+\mu$
& \raggedright growth with reproduction, concave $\phi$: $P'(w)-\eta\,\dot c/c=r+\mu$ \tabularnewline
\raggedright {\scriptsize Reinforce\-ment learning}
& \raggedright $\mathbb{E}\sum_j\gamma^j F_j$  Bellman equation
& \raggedright $-\ln\gamma$; or a per-step hazard
& \raggedright explore vs.\ exploit: $\ln(1+D/\Delta)=-\ln\gamma$; Gittins index $\nu$
& \raggedright growing-stock MDP: $\ln A+ $ $ \ln \frac{x_j}{x_{j+1} }=-\ln\gamma$ \tabularnewline
\bottomrule
\end{tabular}
\caption{Intertemporal problems across fields. In every cell, the decision rule equates the effective growth rate of the reward, $\gamma_R$ (left side), with the discount rate, $\gamma_G$ (right side). Approaches differ mainly in the origin of $\gamma_G$: a preference parameter in psychology, economics and reinforcement learning, but endogenous (population growth plus mortality) in life history theory. All symbols are summarized in Appendix~\ref{app:notation}.}
\label{tab:traditions}
\end{table}

\section{Point problems: optimal waiting times and discount rates}
\label{sec:point examples}

\subsection{Exponential and linear rewards}
The simplest case has constant reward and discount rates: $R(\tau,t)=R_0\,e^{\gamma_R(\tau-t)}$ and $\gamma_G$. If $\gamma_R>\gamma_G$, the agent should wait (prefer the ``larger later'' option, LL); if $\gamma_R<\gamma_G$, it should collect immediately (``smaller sooner'', SS). The indifference point, $\gamma_R=\gamma_G$, does not depend on time  so there are no preference reversals: an agent that starts out waiting waits forever, and there is no optimal waiting time.

\begin{figure}[htbp]
    \centering
    \includegraphics[width=1\textwidth]{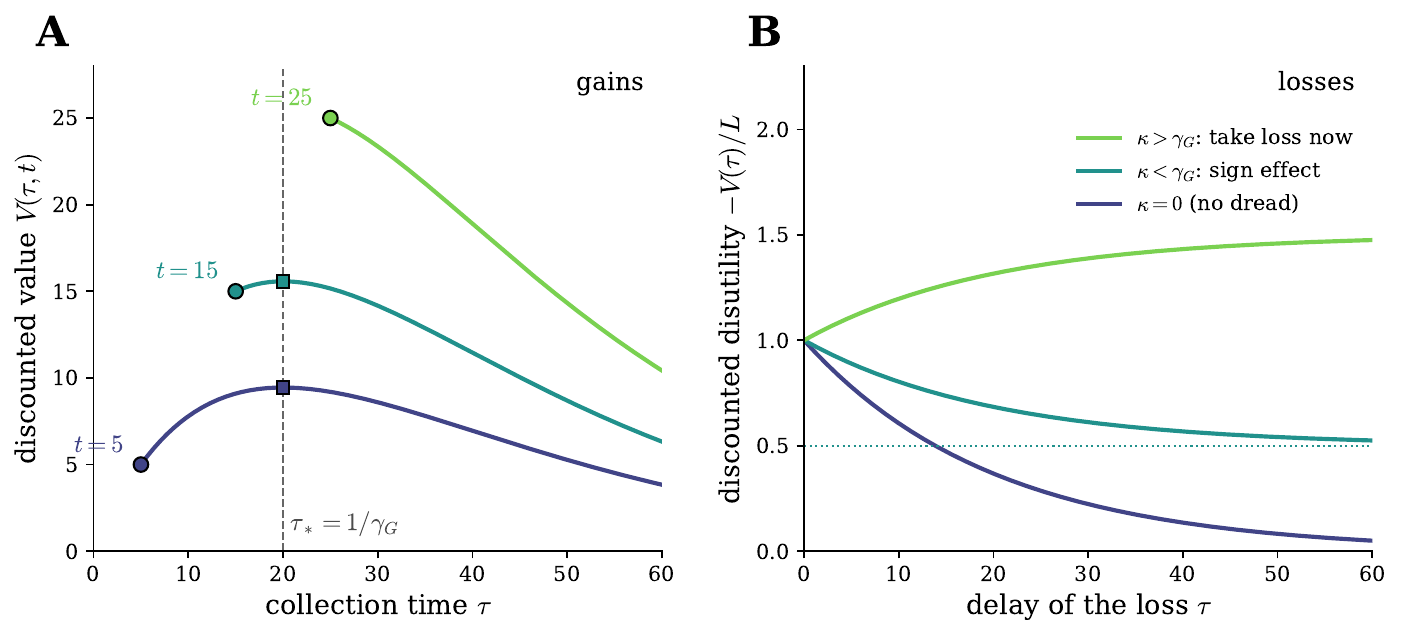}
    \caption{(A) Gains: the discounted value $V(\tau,t)=R_0\,\tau\,e^{-\gamma_G(\tau-t)}$ of a linearly growing reward collected at $\tau\geq t$, for $t \in \{5,15,25\}$ (circles), $R_0 =1$ and $\gamma_G=0.05$. Value first rises as the reward grows and then falls as discounting dominates. The optimal collection time is $\tau_* = 1/\gamma_G = 20$ (dashed line, squares) for $t = 5$ and $t=15$; a reward available at $t = 25$ or later is best collected immediately. (B) Losses: the discounted disutility of a loss of fixed size $L$ due at delay $\tau$, with dread cost $\kappa L$ per unit time while waiting, $-V/L=G_{\rm eff}(\tau)$, Eq.~\eqref{eq:dread}, for $\gamma_G=0.05$. Without dread ($\kappa=0$) losses are discounted like gains. With mild dread ($\kappa=0.025<\gamma_G$) disutility levels off at $\kappa/\gamma_G$ (dotted line), so losses appear discounted less steeply than gains (the sign effect). With strong dread ($\kappa=0.075>\gamma_G$) delay increases disutility and the loss is best taken immediately.}
    \label{fig:optimal waiting time}
\end{figure}

A preference reversal and an optimal waiting time require that the two rates change differently over time. For example, consider a reward that grows linearly, $R(\tau)=R_0\,\tau$ (e.g.\ a resource accumulating at a constant rate), with constant exponential discounting rate $\gamma_G$. Then
\begin{eqnarray}
    V(\tau,t) = R_0\, \tau\, e^{-\gamma_G (\tau-t)};  \qquad  \frac{\partial \ln V}{\partial \tau} = \frac{1}{\tau} - \gamma_G.
\end{eqnarray}
Value rises at first and is eventually overtaken by discounting (figure~\ref{fig:optimal waiting time}A). The optimal collection time is $\tau_*=1/\gamma_G$, starting at  any $t<1/\gamma_G$. For $t\geq1/\gamma_G$, the reward should be collected immediately. The agent therefore waits (LL) early in life and switches to SS at $\tau_*$. In a laboratory choice between collecting at $t$ or at $\tau>t$, the agent prefers LL if and only if
\begin{equation}
\frac{R(\tau)}{R(t)} > \exp\left(\int_t^\tau \gamma_G(u)\,du\right)
\label{eq:SS-LL}
\end{equation}
(see Table~\ref{tab:linear accumulation} in Appendix~\ref{app:SSLL} for explicit cases). Because the reward grows, a smaller $\gamma_G$ yields a larger reward at $\tau_*$. If discounting is psychological, the agent does well to reduce it; if it reflects environmental hazards, it may be beyond the agent's control, requiring policy or institutional intervention.

\subsection{Losses, dread and savoring}
The same mathematics applies to losses, but the behavioral consequences differ. Since a loss is a negative reward,  the agent now minimizes the discounted loss $L\,G$, with $\partial_\tau\ln(L\,G)=\gamma_L-\gamma_G$ and $\gamma_L=\partial_\tau\ln L$. The condition $\gamma_L=\gamma_G$ again defines $\tau_*$, now a minimum of the discounted loss. For example, $L=L_0\,e^{\lambda\tau^2/2}$ with constant $\gamma_G$ gives $\gamma_L=\lambda\tau$ and $\tau_*=\gamma_G/\lambda$. The agent should postpone a loss while it grows more slowly than discounting shrinks it, $\gamma_L<\gamma_G$. This means that an optimal waiting time requires accelerating losses, whereas for gains it requires decelerating rewards. If $\gamma_L<\gamma_G$ at all times, a mounting loss is postponed forever, a seemingly `rational' but ruinous course familiar from runaway personal or national debt.

Two behavioral effects are well documented. Many people expedite losses, preferring a larger penalty sooner to a smaller one later in order to `get it over with'~\cite{loewenstein1987,berns2006}. Also, losses are often discounted less steeply than gains of the same size: This is known as the ``sign effect''~\cite{thaler1981,loewenstein1992,frederick2002}. Both effects can be explained in terms of anticipation (``dread''), which adds to the loss itself~\cite{loewenstein1987}. To see this, let a loss of constant magnitude $L$, due at $\tau$, carry a dread cost $\kappa L$ per unit time while waiting. With exponential discounting (and $t=0$), the total discounted disutility is
\begin{align}
    -V(\tau) &= L\, G(\tau) + \int_0^\tau \kappa L\,G(s)\,ds = L\, G_{\rm eff}(\tau), \qquad
    G_{\rm eff }(\tau) = e^{-\gamma_G\tau} + \frac{\kappa}{\gamma_G}\big(1 - e^{-\gamma_G \tau}\big),
    \label{eq:dread}
\end{align}
so the effective discount rate at short delays is $\gamma_G-\kappa$. When $0<\kappa<\gamma_G$, $G_{\rm eff}$ decays more slowly than $G$ and levels off at $\kappa/\gamma_G$: losses appear to be discounted less steeply (the sign effect; figure~\ref{fig:optimal waiting time}B). When $\kappa>\gamma_G$, the effective discount rate is negative, delay increases total disutility, and the agent takes the loss immediately, $\tau_*=0$. Consistently,  in experimental settings, individuals whose neural responses during waiting rise more steeply are those who choose larger shocks sooner~\cite{berns2006}. Positive anticipation of gains (``savoring'') acts symmetrically. By  lowering the effective discounting of future rewards, savoring can lead people to `save the best for last'~\cite{loewenstein1987}. 

\subsection{Uncertainty, risk aversion and hyperbolic discounting}
\label{sec:uncertainty}
Uncertainty is a first step towards deriving $R$ and $G$ from underlying processes. It acts differently depending on whether it affects rewards or discounting.

\noindent{\bf Uncertain rewards.} Let resources $y$ grow by geometric Brownian motion, the standard model of stochastic growth in population dynamics and finance~\cite{merton1969},
\begin{equation}
dy = g\, y\, dt + \sigma\, y\, dB,
\end{equation}
with mean growth rate $g$, volatility $\sigma$ and Brownian increment $dB$. Then $y(\tau)=y_0\exp[(g-\sigma^2/2)\tau+\sigma B_\tau]$ is lognormal. With power law (constant relative risk aversion, CRRA) utility, $u(y)=y^{1-\eta}/(1-\eta)$, where $\eta\geq0$ is relative risk aversion~\cite{pratt1964} (Appendix~\ref{app:RRA}),
$\mathbb{E}[y^{1-\eta}] = y_0^{1-\eta}\exp[(1-\eta)(g - \eta\sigma^2/2)\tau]$. Taking the reward to be the certainty equivalent, ${\rm CE}=u^{-1}(\mathbb{E}[u(y)])$,
\begin{equation}
R(\tau) = y_0\, e^{(g - \eta \sigma^2/2)\tau}, \qquad \text{so that the decision boundary is} \qquad g - \eta \frac{\sigma^2}{2} = \gamma_G.
\end{equation}
We see that volatility and risk aversion combine into a penalty on the growth rate of rewards, the same term that enters precautionary corrections to the social discount rate~\cite{gollier2002}. Greater volatility makes the agent wait less and accept smaller rewards. If the mean growth rate declines, e.g.\ $g(\tau)=g_0\,e^{-\tau/\tau_0}$, there is an optimal waiting time, $\tau_*=\tau_0\ln(g_0/\gamma_{\rm eff})$, with $\gamma_{\rm eff}=\gamma_G+\eta\sigma^2/2$. We see that declines in the growth of rewards are equivalent to increases in the discount rate. With $\gamma_G=0$, discounting arises purely from risk aversion in the face of environmental variability.

The Stone--Geary utility~\cite{stone1954,geary1950}, $u(y)=(y-\bar y)^{1-\eta}/(1-\eta)$, adds a subsistence level $\bar y$, with relative risk aversion $\eta\,y/(y-\bar y)$. It reduces to $\eta$ for a rich agent, $y\gg\bar y$, but diverges as $y\to\bar y$. Thus,  near subsistence, effective discounting becomes extremely sensitive to uncertainty~\cite{decourson2023}.

\noindent{\bf Heterogeneous hazards.} Suppose instead that the discount rate is itself uncertain or composed of many hazards, with a Gamma distribution of mean $\bar\gamma_G$ and shape $n$. Averaging gives
\begin{equation}
G(\tau) = \mathbb{E}\big[e^{-\gamma \tau}\big] =   \Big(1 + \frac{\bar\gamma_G}{n}\,\tau\Big)^{-n}, \qquad
\gamma_G (\tau) = \frac{\bar\gamma_G}{1 + \bar\gamma_G\,\tau/n},
\label{eq:hyperbolic-G}
\end{equation}
which results in a discount rate that declines with delay.  Thus,  an initially impatient agent reverses its attitude and prefers waiting at long delays. Equation~\eqref{eq:hyperbolic-G} is the generalized hyperbola of Loewenstein and Prelec~\cite{loewenstein1992}, and its case $n=1$, $G=1/(1+\bar\gamma_G\tau)$, is the hyperbolic function widely used in psychology~\cite{ainslie1975,mazur1987}; exponential discounting is recovered as $n\to\infty$. The same averaging underlies normative derivations of hyperbolic discounting from uncertain hazards in biology and psychology~\cite{sozou1998,story2016} and of declining social discount rates in economics~\cite{weitzman2001,dasgupta2005}.

In summary, uncertainty in the growth of an asset reduces effective rewards and acts as discounting, whereas a wide distribution of hazards makes the discount rate decline in time, so that long waits are discounted less. These are both common patterns of human behavior~\cite{frederick2002,cohen2020}.

\subsection{Exploration as investment: multi-armed bandits}
\label{sec:RL}
Reinforcement learning (RL) studies agents that learn to act so as to maximize discounted rewards~\cite{sutton2018} (Appendix~\ref{app:RL}). Its discount function is $G(j)=\gamma^j$ per step $j$, so that $\gamma_G=-\ln\gamma$. A classic problem is the multi-armed bandit, in which an agent must discover the payoffs of several arms by pulling them. Consider a safe arm that pays $F_s$ per pull, $0<F_s<1$, and a risky arm that pays $1$ if good and $0$ if bad, with prior probability $\psi<F_s$ of being good; one pull reveals its type. The choice to explore is a point problem in which the reward is information. Pulling the risky arm once costs $\Delta=F_s-\psi$ in expected immediate payoff. In exchange, with probability $\psi$ the agent finds a better arm and gains $1-F_s$ on every later pull: the information acts as an asset paying an expected dividend $D=\psi(1-F_s)$ per step. One unit invested in exploration therefore returns $D/\Delta$ per step, so its reward grows at the rate
\begin{equation}
\gamma_R=\ln\Big(1+\frac{D}{\Delta}\Big)=\ln\Big[1+\frac{\psi(1-F_s)}{F_s-\psi}\Big],
\end{equation}
and exploration pays, $\sum_{j\geq1}\gamma^j D>\Delta$, when $\gamma_R>\gamma_G$. Equivalently, the agent explores if
\begin{equation}
    F_s < \nu \equiv \frac{\psi}{1 - \gamma + \gamma \psi},
\end{equation}
where $\nu$ is the risky arm's Gittins index~\cite{gittins1974,gittins1979}, the safe payoff at which the agent is indifferent. A myopic agent ($\gamma\to0$, $\gamma_G\to\infty$) has $\nu=\psi$ and compares only immediate payoffs. A patient agent ($\gamma\to1$, $\gamma_G\to0$) has $\nu\to1$ and always explores. For any $\gamma>0$, $\nu>\psi$: the agent explores even when the risky arm's expected payoff is lower. The premium $\nu-\psi$ is the value of information, which plays the role that asset appreciation plays in other problems. Impatience suppresses exploration and lowers lifetime payoffs, a link also observed in human choice~\cite{daw2006}. With noisy outcomes, the decision recurs every step and becomes an optimal stopping problem.

\subsection{Life-course rewards: when to retire}
\begin{figure}[htbp]
    \centering
    \includegraphics[width=1\textwidth]{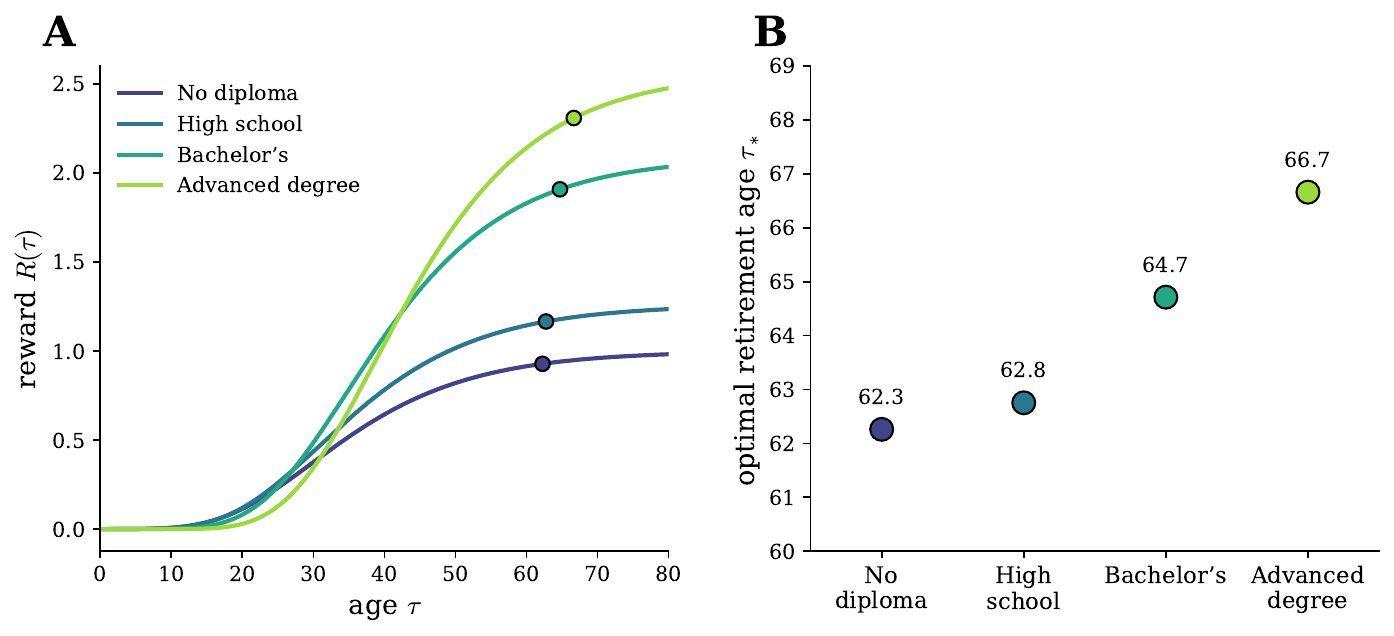}
    \caption{(A) Reward streams rising and plateauing at different ages approximate wages over the life course; $\omega = 0.08$ (shared) and $(\tau_w, R_{\max})$: no diploma $(29.73,\,1.00)$, high school $(30.73,\,1.26)$, bachelor's $(34.73,\,2.09)$, advanced degree $(38.73,\,2.57)$. Circles mark the optimal retirement age on each curve. (B) The optimal collection time $\tau_*$ for each stream, from $\gamma_R(\tau_*)=\gamma_G(\tau_*)$, with Gompertz--Makeham mortality $\mu_0 = 5\times10^{-4}$, $\alpha=2\times 10^{-5}$, $\beta =0.09$.}
    \label{fig:s-shaped-rewards}
\end{figure}

Consider retirement as a point problem: when to stop accumulating a wage stream. We model wages over the life course by an S-shaped (Gompertz) curve,
$R(\tau) = R_{\max}\, e^{-e^{-\omega(\tau-\tau_w)}}$,
near zero in childhood, rising fastest at age $\tau_w$ and approaching $R_{\max}$ (figure~\ref{fig:s-shaped-rewards}A). We let the plateau be higher for a later inflection, $dR_{\max}/d\tau_w>0$, representing sustained investment in skills, as with higher education. Then $\gamma_R(\tau)=\omega\,e^{-\omega(\tau-\tau_w)}$, which declines monotonically with age. We take the discount rate to be the Gompertz--Makeham mortality hazard~\cite{gompertz1825,makeham1860}, $\gamma_G(\tau)=\mu(\tau)=\mu_0+\alpha\,e^{\beta\tau}$, which is small in early adulthood and grows exponentially with age.

A declining $\gamma_R$ and an increasing $\gamma_G$ imply an optimal age at which to take the reward, $\gamma_R(\tau_*)=\mu(\tau_*)$. Reward streams representing higher educational attainment lead to longer patience (figure~\ref{fig:s-shaped-rewards}B), in qualitative agreement with the greater measured patience of higher-income and more educated individuals~\cite{lawrance1991,reimers2009}. Differences in observed patience arise here entirely from differences in reward streams because we assumed (simplistically) identical discounting. With these parameters, optimal retirement falls between ages 62 and 67, close to observed retirement ages in the United States. Mortality alone gives realistic ages only if it is taken lower than period life-table values at these ages (here $\mu(65)\approx0.007$ per year, against roughly $0.01$--$0.016$), suggesting that mortality also varies with wages and education, as empirically observed~\cite{preston1975changing,olshansky2012differences,chetty2016association}. 

\subsection{Life history theory: maturity and the timing of reproduction}
\label{sec:LH}
Life history theory treats the timing of growth, reproduction and death as strategies shaped by natural selection~\cite{stearns1992,roff2002,charlesworth1994}. It interprets both rewards and discounting in terms of demographic quantities, so no psychological assumptions are needed. Rewards are measured by Fisher's reproductive value at age $\tau$~\cite{fisher1930},
\begin{equation}
v(\tau) = \frac{1}{l(\tau)}\int_\tau^\infty e^{-r(a-\tau)}\,l(a)\,m(a)\,da, \quad  \rightarrow \quad   \int_0^\infty e^{-ra}\,l(a)\,m(a)\, da = 1,
  \label{eq:euler lotka}
\end{equation}
where $l(a)$ is survivorship to age $a$, $\mu(a)=-d\ln l/da$ is age-specific mortality, $m(a)$ is fecundity, and $r$ is the population growth rate. Normalizing $v(0)=1$ gives the Euler--Lotka equation, on the right. Here, the reward stream is fecundity, $F=m(a)$, and the discount factor is $G(a)=e^{-ra}\,l(a)$, with discount rate $\gamma_G=r+\mu(a)$ made up of the population growth rate, which dilutes later offspring, plus age-specific mortality. This is the evolutionary origin of time preference identified by Rogers and others~\cite{rogers1994,hansson1990,sozou2003,chu2010}. Differentiating $v$ gives the value identity (Appendix~\ref{app:LH})
 \begin{equation}
   \frac{d v }{d \tau} = (r+\mu) \,v  - m,
   \label{eq:v identity}
 \end{equation}
which states that the growth rate of reproductive value, $v'/v$, plus the reward yield, $m/v$, equals the discount rate, $r+\mu$, for any life history. 

The optimality principle in evolutionary theory maximizes $r$. Varying the Euler--Lotka equation gives the sensitivities of fitness (Appendix~\ref{app:LH})
\begin{eqnarray}
    \frac{\delta r}{\delta\mu(a)} = -\frac{e^{-ra}\,l(a)\,v(a)}{T}, \qquad \frac{\delta r}{\delta m(a)} = \frac{e^{-ra}\,l(a)}{T},
    \label{eq:hamilton}
\end{eqnarray}
where $T=\int_0^\infty a\,e^{-ra}\,l(a)\,m(a)\,da$ is the generation time. These are Hamilton's forces of selection~\cite{hamilton1966,caswell2001}. Optimality requires instead that $r$ is stationary under variations of the organism's choices, such as the age at maturity, $\tau_M$. Writing the Euler--Lotka integral as $W(r;\tau_M)=1$, with $\partial W/\partial r=-T$, gives $dr/d\tau_M=T^{-1}\,\partial W/\partial\tau_M$. Thus, maximizing $r$ is equivalent to setting the variation of the value integral, at fixed $r$, to zero~\cite{taylor1974,schaffer1974}.

\noindent{\bf Age at maturity.} Let fecundity be zero before maturity and $m(\tau_M)$ afterwards, with constant mortality, $\mu$. The Euler--Lotka equation becomes $1=m(\tau_M)\,e^{-(r+\mu)\tau_M}/(r+\mu)$, so that $R=m(\tau_M)$, $G=e^{-(r+\mu)\tau_M}$, and the optimal age at maturity satisfies
\begin{equation}
\gamma_R = \frac{ 1}{m}\frac{d m}{d \tau_M} = r + \mu = \gamma_G.
\end{equation}
Organisms should delay maturity while fecundity grows faster with age than mortality plus population growth. Higher mortality (harsher environments) or faster population growth shift reproduction to younger ages, the classic trade-off associated with $r$--$K$ selection~\cite{pianka1970,stearns1992,charnov1993}. With juvenile mortality $\mu_J$ different from adult mortality $\mu_A$, the equation becomes $1=m(\tau_M)\,e^{-(r+\mu_J)\tau_M}/(r+\mu_A)$ and the optimum is $m'/m=r+\mu_J$. We see that  only juvenile mortality discounts the wait, and that lower $\mu_J$ may promote delayed maturity.

\noindent{\bf Improvements in fertility or survival: sooner or later?} The forces of selection, Eq.~\eqref{eq:hamilton}, answer the evolutionary analogue of a smaller-sooner versus larger-later choice: at what age is an improvement in fertility, or in survival, most valuable to fitness? Each sensitivity has the point form $V=R\,G$.

\emph{Fertility.} A proportional increase in fecundity at age $\tau$, $\delta m(a)=\varepsilon\,m(\tau)\,\delta(a-\tau)$, gives $\delta r=\varepsilon\,e^{-r\tau}\,l(\tau)\,m(\tau)/T$. The reward is $R=m(\tau)$ and the discount function is $G=e^{-r\tau}l(\tau)$, with $\gamma_G=r+\mu$, so the most valuable age satisfies $m'/m=r+\mu$: the mode of the age distribution of mothers in the stable population. For a human-like life history (Appendix~\ref{app:LH}), this gives $\tau_*\approx26.4$ years, shortly before peak fecundity at $27.2$ years.

\emph{Survival.} A reduction of mortality of fixed size at age $\tau$, $\delta\mu(a)=-\varepsilon\,\delta(a-\tau)$, gives $\delta r=\varepsilon\,e^{-r\tau}l(\tau)v(\tau)/T$. Now $R=1$ and $G=e^{-r\tau}l\,v=\int_\tau^\infty e^{-ra}l\,m\,da$ is the reproduction still to come, whose discount rate is $\gamma_G=m/v\geq0$ (Appendix~\ref{app:LH}). Since the reward does not grow, a survival improvement is always (weakly) preferred sooner, and is worthless after reproduction ends. This is Hamilton's declining force of selection with age~\cite{hamilton1966}, and mirrors the empirical finding that health gains are valued earlier in life~\cite{odum2020}. 

\section{Path problems: sustained intertemporal optimal behavior}
\label{sec:path examples}

Path problems are usually treated with optimal control or dynamic programming~\cite{kamien1991,bellman1957}. Here we take a short cut to emphasize the structure of rewards and appreciation from Section~\ref{sec:path}. We begin with life history, where a single choice becomes a trajectory, and then show the same structure in economics.

\subsection{Growth versus reproduction}
The classic life-history trade-off is between growth and reproduction when fecundity increases with body size~\cite{kozlowski1992,perrin1993,roff2002}. Let an organism of mass $w$ with net energy production $P(w)$ split its energy between reproduction, $c$, and growth,
\begin{equation}
    \dot w = P(w) - c,
\end{equation}
with fecundity $m=\phi(c)$ and constant mortality $\mu$. Fitness follows from the Euler--Lotka equation,
\begin{equation*}
1=\int_0^\infty e^{-(r+\mu)a}\,\phi[c(a)]\,da.
\end{equation*}
One unit of energy withheld from reproduction at age $t$ and invested in growth compounds into $\Phi(\tau,t)$ units at age $\tau$, Eq.~\eqref{eq:Phi} with $h'=P'(w)$, giving
\begin{equation}
    R(\tau,t) = \phi'\big[ c(\tau)\big]\,\Phi(\tau,t), \qquad G(\tau,t) = e^{-(r+\mu)(\tau-t)},
\end{equation}
and the condition $\gamma_R=\gamma_G$, Eq.~\eqref{eq:path condition},
\begin{equation}
P'(w) - \eta\,\frac{\dot c}{c} = r + \mu, \qquad \eta \equiv -\frac{c\,\phi''(c)}{\phi'(c)}.
\label{eq:LH path}
\end{equation}
The curvature $\eta$ plays the same role as relative risk aversion in Section~\ref{sec:uncertainty}. For $\eta=0$, Eq.~\eqref{eq:LH path} is a condition on the state alone and the solution is a point; otherwise it is a trajectory for $c$.

If fecundity is proportional to energy allocated, $\phi=\phi_0c$, then $\eta=0$: all energy goes to growth until a size $w_*$ at which $P'(w_*)=r+\mu$, and to reproduction thereafter, a bang-bang solution corresponding to determinate growth~\cite{kozlowski1992,perrin1993}. For the ontogenetic growth model of West \emph{et al.}~\cite{west2001}, $P(w)=a_0w^{3/4}-b_0w$, this gives
\begin{equation}
    w_* = \Big[\frac{3a_0}{4(b_0 + r + \mu)}\Big]^4,
\end{equation}
so that higher discounting (higher mortality or faster population growth) results in a smaller adult size. If instead fecundity has diminishing returns, $\phi''<0$ (for example because larger clutches lower offspring survival), we obtain a path problem,
\begin{equation}
    \frac{\dot c}{c} = \frac{P'(w) - r - \mu}{\eta},
\end{equation}
together with $\dot w=P(w)-c$. The organism grows and reproduces over time, and reproductive investment grows more slowly the higher the discounting, converging to a maximal size as growth slows. An S-shaped $\phi$ combines both regimes towards more realistic behavior: First the organism grows, then there is a jump into reproduction at a threshold size, followed by a path with declining growth and reproduction, as in the concave case.

\subsection{Resource extraction and capital accumulation in economics}
Many intertemporal problems in economics maximize discounted utility~\cite{ramsey1928,samuelson1937}, defined as
\begin{equation}
W = \int_0^\infty  u[c(t)] \, e^{-\gamma_G t}\,dt,
\label{intertemporal-utility}
\end{equation}
where $u$ is the utility stream, $c(t)$ is consumption at time $t$, and $\gamma_G$ is the pure rate of time preference (often written $\rho$), subject to the dynamics of a stock that plays the role of body mass in the previous example. We will consider two classic models.

\noindent{\bf Hotelling's rule.} An owner of a non-renewable resource with initial stock $S_0$ extracts at rate $q$, $\dot S=-q$, and earns rent $p-c_e$ per unit (price minus extraction cost), maximizing $\int_0^\infty e^{-\gamma_Gt}[p(t)-c_e]\,q(t)\,dt$~\cite{hotelling1931}. Here $h=0$: a unit in the ground neither grows nor perishes, $\Phi=1$, and each unit has the point structure $V=R\,G$ with $R=p-c_e$. If the rent grows faster than $\gamma_G$, owners wait; if more slowly, they sell everything now. Because payoffs are linear in $q$, each owner would sell at the peak of $V$, but if all did so prices would collapse. Market equilibrium therefore requires $V$ to be flat, giving Hotelling's rule: the rent of an exhaustible resource rises optimally at the discount rate, $\dot R/R=\gamma_G$~\cite{hotelling1931}. With a demand curve $p(q)$, this gives the extraction schedule $\dot q=\gamma_G\,(p-c_e)/(dp/dq)\leq0$; for linear demand, $p=a-bq$ and $c_e=0$, $p(t)=p_0e^{\gamma_Gt}$, where the stock fixes $p_0$ through the exhaustion time. The rule fixes the growth rate of the rent and the stock fixes its level.

\noindent{\bf Capital accumulation.} In the Ramsey--Cass--Koopmans (RCK) model~\cite{ramsey1928,cass1965,koopmans1965}, Eq.~\eqref{intertemporal-utility} is maximized subject to
\begin{equation}
\dot k = f(k) - \delta_k\, k - c,
\end{equation}
where $f(k)$ is the productive output of capital (revenue) and $\delta_k$ accounts for depreciation and dilution by population growth, so that $h=f-\delta_kk$. Unlike Hotelling's problem, the stock appreciates and utility is nonlinear. Withholding a unit of consumption at $t$ and consuming it at $\tau$ gives (Appendix~\ref{app:RCK})
\begin{align}
R(\tau,t) &= u'\big[ c(\tau) \big]\, \exp\int_t^{\tau}\big[f'(k(u))-\delta_k\big]\,du, \qquad G(\tau,t) = e^{-\gamma_G(\tau-t)}.
\end{align}
With CRRA utility, $\gamma_R=f'(k)-\delta_k-\eta\,\dot c/c$. With linear utility ($\eta=0$) this would again be a point problem. But with concave utility, remaining on $\gamma_R=\gamma_G$ gives the Keynes--Ramsey rule~\cite{ramsey1928,cass1965},
\begin{equation}
\frac{\dot c}{c} = \frac{f'(k) - \delta_k - \gamma_G}{\eta},
\label{eq:Keynes-Ramsey}
\end{equation}
which, with the capital equation and a transversality condition, determines the saddle path for $(k,c)$. At the steady state, $f'(k_*)=\gamma_G+\delta_k$: the physical return of waiting equals the full discount rate, and depreciation acts as discounting. Equation~\eqref{eq:Keynes-Ramsey} is the same as the life-history condition, Eq.~\eqref{eq:LH path}, with capital in place of body mass. It also shows why the consumption discount rate used in policy, $\gamma_G+\eta\,\dot c/c$, differs from the pure rate of time preference~\cite{heal2007}.

\noindent{\bf Subsistence and the impatience of the poor.} Steep discounting is often associated with poverty~\cite{lawrance1991,haushofer2014,mell2021,pepper2017}. With Stone--Geary utility, $u(c)=(c-\bar c)^{1-\eta}/(1-\eta)$, marginal utility $u'=(c-\bar c)^{-\eta}$ gives
\begin{equation}
  \frac{\dot c}{c} = \frac{f'(k) - \delta_k - \gamma_G}{\eta} \,\frac{c - \bar c}{c}, \qquad
  f'(k)-\delta_k = \gamma_G + \eta\,\frac{c}{c - \bar c}\,g_c,
\end{equation}
where $g_c=\dot c/c$. As $c\to\bar c$, consumption growth vanishes: close to subsistence the agent can barely save. Moreover, the return on capital required to justify any given consumption growth must become very large. The resulting present orientation is not caused by a higher $\gamma_G$, but by the growing inability to forgo consumption as the subsistence threshold approaches, in line with recent threshold-based models of poverty and discounting~\cite{decourson2023,mell2021}. Measured with standard methods, such an agent would nonetheless appear to have a (very) high discount rate~\cite{lawrance1991,cohen2020}.

\section{Discussion}
\label{sec:discussion}
Weighing costs and rewards over time is at the heart of the most consequential problems in nature and society, from reproductive strategies underlying evolution, to dilemmas of cooperation, to learning, innovation and economic growth.  In this paper, we set out to ask whether intertemporal decisions studied separately in evolution, ecology, psychology, economics and machine learning share a common structure. We have shown that they do and that their underlying logic can be made simple and transparent. In every case, the optimal decision compares two temporal rates: the growth rate of the reward of one unit of a good or asset held, $\gamma_R$, to  the discount rate, $\gamma_G$. For a single decision, their equality determines the existence and value of an optimal waiting time.  For continuous allocation problems, this condition becomes an equation of motion such as the Keynes--Ramsey rule in economics, the growth--reproduction schedule in life history theory, or the Bellman condition in reinforcement learning. The appreciation factor $\Phi$ connects the two cases, by factoring into the decision the amplification of the asset held over time.  Tables~\ref{tab:point-path} and~\ref{tab:traditions} summarize this correspondence. 

The comparative formal view developed in this paper highlights several points that are, in our view, obscured when each field is taken in isolation. First, 
discount  rates are always relative, not absolute. The decision to wait depends on $\gamma_R-\gamma_G$, not on $\gamma_G$ alone. This means that there is an invariance of the difference of rates: each rate can be shifted if compensated by the other. In particular, a slowing down of reward growth is observationally equivalent to faster discounting (Section~\ref{sec:uncertainty}). Measuring only discount rates therefore mixes the agent's discounting with its expectations about how rewards evolve: their growth, volatility, perishability and the curvature of their value functions (utility, fertility). This may help explain why discount rates vary with commodity~\cite{odum2020} and magnitude~\cite{thaler1981,loewenstein1992}, are hard to identify experimentally, and predict real-world behavior poorly~\cite{cohen2020,lipman2024}. Food, which perishes quickly, has a negative appreciation rate, $h'<0$, and should appear more steeply discounted than money, which is more stable and can earn interest, even with identical $\gamma_G$. Outside the laboratory, where rewards cannot be easily determined, explaining and changing intertemporal behavior requires as much attention to reward streams as to discounting.

Second, discount rates have root causes. Economics and machine learning typically treat their discount rate as an exogenous preference or hyperparameter~\cite{samuelson1937,frederick2002,sutton2018}. Life history theory, by contrast, derives it fully from demography, as mortality hazard plus the population growth rate~\cite{fisher1930,rogers1994,hansson1990}. In some cases, economics has partly endogenized discounting, especially  in terms of mortality rates~\cite{yaari1965,acharya2004}, and impatience may depend on the agent's state or investments~\cite{obstfeld1990,hayakawa1997,becker1997}, although whether this is consistent with observed consumption patterns is debated~\cite{hayakawa2019}. Depreciation enters the effective discount rate, $\gamma_G+\delta_k$, and the consumption discount rate depends on the growth and curvature of associated utility functions~\cite{heal2007}. The examples of Sections~\ref{sec:point examples} and~\ref{sec:path examples} show how hazards, uncertainty, background growth and depreciation each produce (additive) components of the effective discount rate. This suggests a resolution of the long-standing finding that mortality is far too weak to explain observed impatience~\cite{mell2021,stephens2004,henly2008}.  In our framework, mortality is only one contribution to $\gamma_G$. The opportunity costs of waiting~\cite{fawcett2012,mell2021}, volatility combined with risk aversion, and proximity to subsistence thresholds all act through $\gamma_R$ with effects that can be much larger, at least for shorter temporal scales.

Third,  a catalogue of candidate mechanisms follows from this sort of comparative analysis. These include hazards, including mortality, interruption and perishability. Also, when hazards are heterogeneous or uncertain, their average effects produce the declining discount rates of hyperbolic discounting~\cite{sozou1998,story2016,weitzman2001}. They also include uncertainty in rewards, which reduces their effective growth in proportion to risk aversion. Subsistence thresholds are likely crucial in situations where agents are very deprived or desperate, leading to risk aversion, exceptionally present-oriented behavior and a search for improbably high-yield strategies, such as lotteries~\cite{decourson2023}. Depreciation and dilution are also important and include inflation, capital depreciation and population growth, and more generally any change in the standard against which rewards are evaluated, including an agent's relative position among competitors.  Psychological flows during waiting may also play a role, such as dread and savoring, which change effective growth rates for losses and gains~\cite{loewenstein1987,berns2006}.  The value of information is important in long-horizon problems where learning is possible, making agents explore more rather than immediately exploit. Each of these mechanisms makes present-oriented behavior the rational response to an agent's circumstances, not a deficit of character or a subjective preference~\cite{kidd2013,story2016,daly2005,pepper2017}. In the same vein, the U-shaped profile of time preference over life found in meta-analyses of human data~\cite{lu2023} and in life-history models~\cite{sozou2003,chu2010,kageyama2011} arises naturally when reward growth declines with age while mortality hazards rise, as in our retirement and maturity examples. In real-world situations, therefore, it is important to test which of these types of effects may be at play, and to what extent one dominates the magnitude of discounting or if several contribute (additively) to produce observed behavior. 

Our treatment  of intertemporal discounting problems was deliberately minimal for the sake of clarity of comparative analysis and, therefore, has a number of limitations. For example, most of our examples are deterministic or use simple stochastic processes.  We considered only a single agent, and took the discount function as given in most cases. We did not derive present bias of the quasi-hyperbolic kind~\cite{strotz1955,laibson1997}. We also ignored the effects of aggregate (correlated) risk on the evolution of time preference~\cite{robson2009}, or the foreground--background structure of natural foraging choices that can make impulsive rules adaptive without discounting~\cite{stephens2004,fawcett2012}. All of these situations can be written in the language of $R$ and $G$, but require their own elaborations.

The framework developed here points to several interesting open problems. Experiments and field studies should explicitly manipulate and measure reward dynamics (growth, volatility and perishability) separately from hazards, so that $\gamma_R$ and $\gamma_G$ can be separately identified rather than estimated as a single discount rate~\cite{cohen2020,lipman2024}. Social and strategic hazards, such as competition, conflict and the actions of others, are major sources of uncertainty for real agents, yet they are rarely modeled explicitly in any of the traditions reviewed here. Relative social position may help explain the steep discounting associated with low socioeconomic status~\cite{haushofer2014,pepper2017}. Also, real agents must estimate their hazards and reward streams from experience. How they do so and how errors in these estimates propagate to behavior~\cite{story2016,sozou2003}, and how learning agents may discount adaptively~\cite{fedus2019,kurthnelson2009} remain open questions for neuroscience and for artificial agents acting over long horizons. Finally, if steep discounting reflects reward streams and hazards rather than fixed preferences or psychological traits, interventions that permit and raise returns to waiting such as living wages, various forms of insurance, access to credit, reduced volatility and better information, should increase patience and help people escape poverty traps~\cite{haushofer2014,mell2021}. The same logic also bears on collective problems that span generations, such as climate change, where the choice of discount rate dominates policy~\cite{stern2007,nordhaus2007,weitzman2001,gollier2002}. Thus, a clearer and more comprehensive understanding of how discounting and reward streams are perceived and accessed across domains should help design environments in which  patience is affordable in the short term and pays off over a lifetime, and where the beneficial risk-taking behind learning and innovation becomes more universally accessible.

\begin{acknowledgments}
We thank Jordan Kemp, Andrew Stier, and Marc Berman for discussions and members of the Santa Fe Institute working group on the Psychology of Human Development, whose discussions sparked the initial questions leading to this work. This research was partially supported by the Paglia Post-Baccalaureate Research Fellowship from Carleton College (to ET), and by the Conard Initiative on the Foundations of Complexity in Nature and Society (to LMAB).

\noindent\textbf{Author contributions.} E.T.: conceptualization, formal analysis, visualization, writing; L.M.A.B.: conceptualization, formal analysis, supervision, writing. All authors gave final approval for publication.

\noindent\textbf{Data availability.} This article has no additional data. Code used to produce the figures is available at \url{https://github.com/Urban-Science-Lab/temporal_discounting}.

\noindent\textbf{Competing interests.} We declare we have no competing interests.
\end{acknowledgments}

\appendix

\section{Choices between sooner and later rewards}
\label{app:SSLL}
\begin{table}[!htbp]
\centering
\renewcommand{\arraystretch}{1.3}
\begin{tabular}{lcccccc}
\toprule
\textbf{Scenario} & $t$ & $\tau$ & $\tau-t$ & $R(\tau)/R(t)$ & $\exp\left(\int_{t}^\tau \gamma_G(u)\,du\right)$ & \textbf{Decision} \\
\midrule
Baseline window          &  5 & 25 & 20 &  5.000 &  2.718 & LL \\
$\tau$ pushed forward       &  5 & 40 & 35 &  8.000 &  5.755 & LL \\
   &  5 & 60 & 55 & 12.000 & 15.643 & SS \\
$t$ pushed forward       & 15 & 20 &  5 &  1.333 &  1.284 & LL \\
       & 20 & 25 &  5 &  1.250 &  1.284 & SS \\
 & 25 & 30 &  5 &  1.200 &  1.284 & SS \\
\bottomrule
\end{tabular}
\caption{Choices between collecting a reward now (at $t$, SS) or later (at $\tau$, LL), Eq.~\eqref{eq:SS-LL}, for a linearly growing reward, $R(\tau) = R_0\, \tau$, and exponential discounting at the constant rate $\gamma_G = 0.05$, so that $\tau_*=1/\gamma_G=20$. The subject chooses LL when $R(\tau)/R(t)$ exceeds the discount factor.}
\label{tab:linear accumulation}
\end{table}

\section{Relative risk aversion and certainty equivalent}
\label{app:RRA}
\noindent {\bf Relative risk aversion}~\cite{pratt1964} is the negative elasticity of marginal utility with respect to resources, $y$: the percentage decrease in marginal utility per percentage increase in resources. For CRRA utility, $u(y) = y^{1-\eta}/(1-\eta)$, $u'(y)=y^{-\eta}$ and
\begin{equation}
    {\rm RRA} = - \frac{d \ln u'(y) }{d \ln y} = \eta.
\end{equation}
If marginal utility drops rapidly as resources grow, the agent is very averse to losing a fraction of its resources. The same parameter governs how marginal utility changes over time when consumption grows, $d\ln u'/dt=-\eta\,\dot c/c$. This is why $\eta$ appears both in the cost of uncertainty, Section~\ref{sec:uncertainty}, and in the path conditions of Section~\ref{sec:path examples}, where it measures the curvature of utility, or of fecundity $\phi(c)$, as $\eta=-c\,F''/F'$.

\noindent {\bf Certainty equivalent.} For a gamble on a random variable $y$, the certainty equivalent (CE) is the certain amount with the same utility, ${\rm CE}=u^{-1}(\mathbb{E}[u(y)])$. For CRRA utility, ${\rm CE}=(\mathbb{E}[y^{1-\eta}])^{1/(1-\eta)}$. For lognormal $y$ with $\ln y \sim N(m_y, s^2)$, ${\rm CE}=\exp[m_y+(1-\eta)s^2/2]$, which equals the mean for $\eta=0$ and the median for $\eta=1$.

\section{The appreciation factor and feasible variations}
\label{app:Phi}
Here we show that the appreciation factor $\Phi$ of Eq.~\eqref{eq:Phi} expresses the dynamical constraint, Eq.~\eqref{eq:stock dynamics}, and that the two-pulse variations of Eq.~\eqref{eq:two-pulse} generate all feasible variations of a path.

A variation $\delta x$ of the amount cashed in produces a variation $\delta k$ of the stock. Linearizing $\dot k = h(k)-x$ around the optimal path gives
\begin{equation}
\delta \dot k = h'\big(k(s)\big)\,\delta k - \delta x .
\end{equation}
The appreciation factor is the propagator of this equation,
\begin{equation}
\partial_s \Phi(s,t) = h'\big(k(s)\big)\,\Phi(s,t), \qquad \Phi(t,t)=1,
\end{equation}
which gives Eq.~\eqref{eq:Phi}. Dividing the stock variation by $\Phi$,
\begin{equation}
\frac{d}{ds}\left[\frac{\delta k(s)}{\Phi(s,t)}\right] = \frac{\delta\dot k - h'\,\delta k}{\Phi} = -\frac{\delta x(s)}{\Phi(s,t)}.
\end{equation}
The deflated stock, $\delta k/\Phi$, is therefore conserved except when units are cashed in: $\Phi$ converts units at time $s$ into units at time $t$. Integrating from $t$ to $\infty$, with $\delta k(t)=0$ (the stock at $t$ is given) and $\delta k/\Phi\rightarrow 0$ at long times (transversality), gives the budget constraint on feasible variations,
\begin{equation}
\int_t^\infty \frac{\delta x(s)}{\Phi(s,t)}\,ds = 0.
\label{eq:budget}
\end{equation}
Writing any feasible variation as $\varepsilon(s)\equiv \delta x(s)/\Phi(s,t)$, and using $F'[x(s)]\,G(s,t)\,\delta x(s) = V(s,t)\,\varepsilon(s)$ with $V=F'\,\Phi\,G$, the first-order change in value is
\begin{equation}
\delta W = \int_t^\infty V(s,t)\,\varepsilon(s)\,ds, \qquad \int_t^\infty \varepsilon(s)\,ds = 0 .
\end{equation}
With a Lagrange multiplier $\lambda_t$, $\delta W = \int_t^\infty [V(s,t)-\lambda_t]\,\varepsilon(s)\,ds$ must vanish for every $\varepsilon$, so $V(s,t)=\lambda_t$ is constant in $s$: the flatness of $V$ used in Section~\ref{sec:path}. Each two-pulse move is the special case $\varepsilon(u)=\varepsilon\,[\delta(u-s)-\delta(u-t)]$, and every feasible $\varepsilon$ is a superposition of such moves.

For Hotelling's problem, $h=0$, so $\Phi=1$ and Eq.~\eqref{eq:budget} reads $\int \delta q\,ds=0$: units are only moved in time. For linear stock dynamics, $h(k)=\bar r\, k$, the constraint holds exactly: $\int_t^\infty x(s)\,e^{-\bar r (s-t)}ds = k(t)$, the familiar lifetime budget constraint.

\section{Details of the Ramsey--Cass--Koopmans model}
\label{app:RCK}
The reward $R(\tau,t)$ is the utility gain, at time $\tau$, from saving one marginal unit of goods at time $t$: the number of goods that unit becomes, times the utility each yields when consumed. Starting from the optimal path $(c,k)$, consume $\varepsilon$ less at time $t$, add it to capital, and consume the accumulated extra capital at time $\tau$. This returns $k$ to its original path, so the perturbation is feasible.

\noindent {\bf Factor 1: goods grow at the net marginal product.} Linearizing $\dot k = f(k) - \delta_k k - c$ with $c$ fixed gives $d\,\delta k/ds = [f'(k(s)) - \delta_k]\,\delta k$, with $\delta k(t) = \varepsilon$ (here $\delta_k$ is the depreciation rate and $\delta k$ the perturbation of capital), so that
\begin{equation}
\delta k(\tau) = \varepsilon\, \exp\int_{t}^{\tau}\big[f'(k(s)) - \delta_k\big]\,ds = \varepsilon\,\Phi(\tau,t).
\end{equation}

\noindent {\bf Factor 2: each unit is worth its marginal utility.} Consuming the extra goods at $\tau$ raises utility by $u'[c(\tau)]\,\delta k(\tau)$ to first order. Hence
\begin{equation}
R(\tau,t) = u'\big[c(\tau)\big]\, \exp\int_{t}^{\tau}\big[f'(k(s))-\delta_k\big]\,ds, \qquad
\gamma_R = \frac{\partial\ln R}{\partial\tau} = f' - \delta_k - \eta\,\frac{\dot c}{c},
\end{equation}
using $c\,u''/u'=-\eta$. At $\tau=t$, $R=u'[c(t)]$, the utility of consuming the unit immediately. The first term of $\gamma_R$ is the physical return of waiting; the second is the loss from diminishing marginal utility when consumption grows.

\section{Reproductive value and forces of selection}
\label{app:LH}

\noindent {\bf Value identity.}
Write
\begin{equation*}
v(\tau) = \frac{e^{r\tau}\,I(\tau)}{l(\tau)}, \qquad I(\tau)=\int_\tau^\infty e^{-ra}\,l(a)\,m(a)\,da.
\end{equation*}
Since $d\ln[e^{r\tau}/l(\tau)]/d\tau = r+\mu$ and $I' = -e^{-r\tau}\,l(\tau)\,m(\tau)$, we obtain $v' = (r+\mu)\,v - m$, Eq.~\eqref{eq:v identity}. Equivalently, $v(\tau)=\int_\tau^\infty m(a)\,G(a,\tau)\,da$ with
\begin{equation*}
G(a,\tau)=\exp\Big[-\int_\tau^a\big(r+\mu(s)\big)\,ds\Big]:
\end{equation*}
reproductive value is a value integral of the form of Eq.~\eqref{eq:path problem}, with reward flow $m$ and discount rate $r+\mu$.

\noindent {\bf Sensitivity to mortality.}
A variation $\mu\rightarrow\mu+\delta\mu$ changes survival by $\delta l(a) = -l(a)\int_0^a \delta\mu(s)\,ds$. Since the Euler--Lotka equation must still hold, $0 = -T\,\delta r + \int_0^\infty e^{-ra}\,\delta l(a)\,m(a)\,da$. Substituting $\delta l$ and exchanging the order of integration,
\begin{equation}
\delta r = -\frac{1}{T}\int_0^\infty \delta\mu(s)\left[\int_s^\infty e^{-ra}\,l(a)\,m(a)\,da\right]ds
= -\frac{1}{T}\int_0^\infty \delta\mu(s)\,e^{-rs}\,l(s)\,v(s)\,ds,
\end{equation}
the first of Eqs.~\eqref{eq:hamilton}. A uniform reduction, $\delta\mu=-\varepsilon$, gives $\delta r=\varepsilon$ exactly, because
\begin{equation*}
\int_0^\infty e^{-rs}\,l(s)\,v(s)\,ds = T:
\end{equation*}
mortality and population growth enter only through their sum, the discount rate $r+\mu$.

\noindent {\bf Sensitivity to fecundity.}
Since $m$ enters the Euler--Lotka equation linearly, $0=-T\,\delta r+\int_0^\infty e^{-ra}\,l(a)\,\delta m(a)\,da$, which gives the second of Eqs.~\eqref{eq:hamilton}.

\noindent {\bf A human-like life history.} For the example of Section~\ref{sec:LH} we use a Siler mortality curve~\cite{siler1979}, $\mu(a)=\alpha_1e^{-\beta_1a}+\mu_0+\alpha\,e^{\beta a}$, with $\alpha_1=0.15$, $\beta_1=1.2$, $\mu_0=0.005$, $\alpha=3\times10^{-5}$, $\beta=0.10$ (per year), and the fecundity curve $m(a)=m_0\exp[-\vartheta_1(a-a_m)^2+\vartheta_2/(a+1)]$~\cite{colchero2025}, with $m_0=0.1748$, $\vartheta_1=0.015$, $a_m=27$, $\vartheta_2=-5$. Solving the Euler--Lotka equation numerically gives $r\approx0.018$ per year. The density $e^{-ra}l(a)m(a)$ peaks at $\tau_*\approx26.4$ years, where $m'/m=r+\mu$, while $m(a)$ itself peaks at $27.2$ years.

\noindent {\bf Discount rate of survival improvements.}
For a reduction of mortality of fixed size at age $\tau$, $G(\tau)=e^{-r\tau}\,l(\tau)\,v(\tau)=I(\tau)$, so
\begin{equation}
\gamma_G(\tau) = -\frac{d\ln I}{d\tau} = \frac{m(\tau)}{v(\tau)},
\end{equation}
which also follows from Eq.~\eqref{eq:v identity}: $\gamma_G = (r+\mu) - v'/v = m/v$.

\section{Essentials of reinforcement learning}
\label{app:RL}
Reinforcement learning formalizes an agent interacting with an environment over discrete steps, $j=0,1,2,\ldots$, as a Markov decision process~\cite{sutton2018,bellman1957}. At each step the agent observes a state, $k_j$, chooses an action, $x_j$, receives a reward, $F_j=F(x_j)$, and the environment moves to a new state, $k_{j+1}$. The agent seeks the policy that maximizes the expected discounted return,
\begin{equation}
\mathcal{V}(k) = \max\, \mathbb{E}\left[\sum_{j\geq 0}\gamma^j F_j \,\Big|\, k_0 = k\right],
\end{equation}
a discrete version of Eq.~\eqref{eq:path problem} with $G(j)=\gamma^j$ and $\gamma_G=-\ln\gamma$. The discount factor $\gamma\in(0,1)$ is often interpreted as a per-step probability that the interaction continues, so that $1-\gamma$ is a termination hazard, as in life history theory~\cite{sutton2018}. The optimum satisfies the Bellman equation~\cite{bellman1957},
\begin{equation}
\mathcal{V}(k_j) = \max_{x_j}\Big\{F(x_j) + \gamma\,\mathbb{E}\,\mathcal{V}(k_{j+1})\Big\}.
\end{equation}

\noindent {\bf From the Bellman equation to the path condition.}
Let the part of the stock that is not cashed in grow deterministically, $k_{j+1}=h(k_j-x_j)$. The first-order condition is $F'(x_j) = \gamma\,h'(k_j-x_j)\,\mathcal{V}'(k_{j+1})$, and the envelope theorem gives $\mathcal{V}'(k_j) = F'(x_j)$. Shifting one step forward and substituting,
\begin{equation}
F'(x_j) = \gamma\, h'(k_j-x_j)\,F'(x_{j+1}) \quad \Longleftrightarrow \quad \underbrace{\ln h'(k_j-x_j) + \ln\frac{F'(x_{j+1})}{F'(x_j)}}_{\gamma_R} = \underbrace{-\ln\gamma}_{\gamma_G},
\end{equation}
the two-pulse condition of Section~\ref{sec:path} between consecutive steps.

\noindent {\bf A minimal path problem.}
Take $F(x)=\ln x$ and $h(z)=A\,z$. The condition above gives $x_{j+1}/x_j=\gamma A$, and the Bellman equation with $\mathcal{V}(k)={\rm const}+\ln k/(1-\gamma)$ gives the optimal policy $x_j=(1-\gamma)\,k_j$, $k_{j+1}=\gamma A\,k_j$. Consumption rises if $\gamma A>1$ and falls if $\gamma A<1$. With linear reward, $F(x)=x$, the problem becomes a point problem: consume everything now if $\gamma A<1$, never if $\gamma A>1$.

\noindent {\bf Exploration and exploitation.}
In bandit problems the state is the agent's belief about each arm. The value of an arm is summarized by its Gittins index, and for independent arms the optimal policy pulls the arm with the highest index~\cite{gittins1974,gittins1979}. Exploration is an investment: an immediate expected loss buys information, whose return is the stream of better future choices.

\section{Notation}
\label{app:notation}
\begin{table}[!htbp]
\centering
\footnotesize
\renewcommand{\arraystretch}{1.2}
\setlength{\tabcolsep}{3pt}
\begin{tabular}{@{}p{2.66cm}p{5.32cm}p{2.66cm}p{5.32cm}@{}}
\toprule
\textbf{Symbol} & \textbf{Meaning} & \textbf{Symbol} & \textbf{Meaning} \\
\midrule
\multicolumn{4}{@{}l}{\emph{General framework}} \\
$t$ & reference (decision) time & $\tau,\ \tau_*$ & collection time; optimal waiting time \\
$R(\tau,t)$ & reward of one unit collected at $\tau$ & $G(\tau,t)$ & discount function, $G(t,t)=1$ \\
$V=R\,G$ & subjective (discounted) value & $W(t)$ & value integral of a reward stream \\
$\gamma_R,\ \gamma_G$ & growth rate of reward; discount rate & $F[x]$ & reward flow \\
$x$ & amount cashed in per unit time & $k$, $h(k)$ & stock; its growth function, $\dot k=h(k)-x$ \\
$\Phi(\tau,t)$ & units at $\tau$ from one unit withheld at $t$ & $\varepsilon$, $\delta x$, $\delta W$ & size of perturbation; variations \\
\midrule
\multicolumn{4}{@{}l}{\emph{Point problems}} \\
$L,\ \gamma_L$ & magnitude of a loss; its growth rate & $\lambda$, $\kappa$ & loss acceleration; dread intensity \\
$y$, $\bar y$ & resources; subsistence level & $g$, $\sigma$, $B$ & drift; volatility; Brownian motion \\
$\eta$ & relative risk aversion / curvature & $g_0$, $\tau_0$ & initial drift; its decay time \\
$\bar\gamma_G$, $n$ & mean and shape of Gamma-distributed rates & ${\rm CE}$ & certainty equivalent \\
$R_{\max}$, $\tau_w$, $\omega$ & wage plateau, inflection age, rate & $\mu_0$, $\alpha$, $\beta$ & Gompertz--Makeham parameters \\
\midrule
\multicolumn{4}{@{}l}{\emph{Life history theory}} \\
$a$ & age (integration variable) & $r$ & population growth rate (fitness) \\
$l(a)$, $\mu(a)$ & survival function; mortality hazard & $m(a)$ & age-specific fecundity \\
$v(\tau)$ & Fisher's reproductive value & $T$ & generation time \\
$\tau_M$ & age at maturity & $\mu_J$, $\mu_A$ & juvenile, adult mortality \\
$w$, $P(w)$ & body mass; net energy production & $\phi(c)$ & fecundity from reproductive allocation $c$ \\
$a_0$, $b_0$ & ontogenetic growth parameters & $\alpha_1$, $\beta_1$ & Siler juvenile mortality terms \\
$m_0$, $\vartheta_{1,2}$, $a_m$ & fecundity curve parameters & & \\
\midrule
\multicolumn{4}{@{}l}{\emph{Economics and reinforcement learning}} \\
$c$, $u(c)$, $\bar c$ & consumption; utility; subsistence & $f(k)$, $\delta_k$ & output; depreciation rate \\
$g_c$ & growth rate of consumption & $S$, $q$, $p$, $c_e$ & resource stock, extraction, price, cost \\
$a$, $b$ & linear demand parameters & $\gamma$, $F_j$ & per-step discount factor; reward at step $j$ \\
$F_s$, $\psi$, $\nu$ & safe payoff; prior; Gittins index & $\mathcal{V}$, $A$ & RL value function; gross return \\
$\Delta$, $D$ & cost of exploring; information dividend & & \\
\bottomrule
\end{tabular}
\caption{Notation used throughout the paper.}
\label{tab:notation}
\end{table}

\bibliography{refs}

\end{document}